\pdfoutput=1
\documentclass[pdflatex,sn-basic,Numbered,iicol]{sn-jnl}

\usepackage{graphicx}
\usepackage{multirow}
\usepackage{amsmath,amssymb,amsfonts}
\usepackage{booktabs}
\usepackage{array}
\usepackage{longtable}
\usepackage{xcolor}
\usepackage{float}
\usepackage{placeins}

\graphicspath{{./}{./fig/}}

\begin{document}

\title[Managing Shared Hospital Capacity for Medical Tourism]{Managing Shared Hospital Capacity for Medical Tourism: Patient Behavior and Hybrid Simulation}

\author[1,2]{\fnm{Melika} \sur{Baghi}}\email{mbaghi3@gatech.edu}
\author*[1]{\fnm{Hadi} \sur{Mosadegh}}\email{hadi\_mosadegh@aut.ac.ir}

\affil*[1]{\orgdiv{Department of Industrial and Systems Engineering}, \orgname{Amirkabir University of Technology}, \orgaddress{\city{Tehran}, \country{Iran}}}
\affil[2]{\orgdiv{H. Milton Stewart School of Industrial and Systems Engineering}, \orgname{Georgia Institute of Technology}, \orgaddress{\city{Atlanta}, \state{GA}, \country{USA}}}

\abstract{Hospitals that serve medical tourists alongside local patients share beds, specialists, and appointment slots between two groups for whom waiting has different consequences. Operations research models are widely used for hospital capacity and scheduling, but often do not explicitly capture patient behavior; agent-based simulation can represent such behavior alongside discrete-event patient flow, while medical-tourism research still mainly treats traveling patients as cases to be scheduled. We develop a hybrid simulation that represents channel choice, physician switching, treatment abandonment, and medication-related deterioration in a multi-specialty international patient department. A 256-run fractional factorial experiment screens 16 controllable factors across six performance measures, followed by a replicated comparison with a discrete-event-only version without these behaviors. At a bed-constrained configuration, mean inpatient admission waiting time is 42\% lower in the hybrid model (14.8 versus 25.8 days in the discrete-event-only model, $p<0.001$) and 25\% to 64\% lower under two further conditions. Lacking these behaviors, the simpler model shows no abandonment and fewer emergency escalations. The comparison shows how results depend on model structure; neither variant is validated against observed waiting times. A bed-sharing routine in both variants adds no beds as implemented, so bed sharing is not evaluated. The screening identifies bed capacity as the most consistent driver of waiting; priority rules, channel design, and appointment timing also matter, while specialist capacity has less effect on waiting. Managers should review bed capacity and admission rules before adding specialists or tightening schedules.}


\maketitle

\section{Introduction}\label{sec:intro}

Medical tourism creates an important operational challenge for hospitals. Competition among medical-tourism destinations is increasing \citep{Sag2025MedTourismCompetition}, and cross-border care is also being organized through digital channels and specialized intermediaries \citep{Ziemba2025HealthcareGlobalization,Yorulmaz2026Telework}. Patients travel for lower treatment costs, shorter waiting times, or care they perceive as higher in quality \citep{Connell2006,Connell2013,Fetscherin2016}. Once they arrive, the problem becomes one of service delivery. International patients use the same specialists, appointment slots, and inpatient beds as local patients, but a delay costs the two groups differently: a local patient waits for care, whereas a medical tourist also pays for accommodation and a longer trip and may abandon the planned treatment.

Managing this shared system is difficult for two related reasons. The first is how capacity is used. Beds are commonly assigned to clinical sections even when demand is uneven across them, so one section may have patients waiting for admission while another has unused beds; allowing clinically compatible sections to share some beds makes existing capacity more flexible without adding beds. The second is patient behavior. Patients choose between online and in-person consultation, change physicians, leave the system after a poor experience, or need unplanned care when their condition worsens. Each decision changes the demand the hospital sees and, in turn, waiting times, resource use, and the number of patients who complete treatment. The problem is broader than appointment scheduling: managers decide on bed and specialist capacity, channels, admission and priority rules, and appointment timing. These decisions interact, and a useful decision-support model must represent both the hospital process and the patient and resource behaviors that change that process over time.

This study examines whether explicitly representing these behaviors changes the conclusions hospital managers draw. It has three objectives: to assess how patient behavior affects performance, to identify the controllable decisions with the strongest effects, and to translate those findings into practical guidance for a shared local-and-international patient department. To address them, we develop a hybrid simulation of a multi-specialty international patient department\footnote{An interactive version of the model runs in a web browser at \url{https://cloud.anylogic.com/model/ddca6c4e-7a3b-4015-bd9f-fb8d02cb30be}.}, in which agent-based patients and physicians alter the discrete-event patient flow as the system operates and hospital-section agents hold the beds and admission queues. A fractional factorial experiment then screens 16 controllable factors across six performance measures, and a replicated comparison with a discrete-event-only version of the same model shows what changes when the behavioral mechanisms are removed.

In the replicated comparison at a fixed bed-constrained configuration, mean inpatient admission waiting time is about 42\% lower in the hybrid model: 14.8 days against 25.8 days in the discrete-event-only model. The gap persists when demand is raised to the screening level and when the screening's larger bed counts are used. Because several behaviors are removed together, the difference reflects their combined effect in the model as implemented, not the effect of any one of them. The hybrid model also reveals abandonment and adherence-driven emergency escalation that the discrete-event-only version omits, and the screening identifies bed capacity, patient-priority rules, channel design, and the clinic slot interval as the decisions with the strongest effects.

The study models patient behavior in a shared local-and-international patient department, a setting the reviewed hybrid patient-flow literature has not addressed; the behaviors are represented inside the model, not treated as external variability.

Section~\ref{sec:lit} reviews the literature, Section~\ref{sec:problem} describes the problem, Section~\ref{sec:method} the model and experiments, Section~\ref{sec:results} the results, and Section~\ref{sec:managerial} the managerial implications; Section~\ref{sec:conclusion} concludes.

\section{Literature Review}\label{sec:lit}

Three strands of research bear on this problem: hybrid simulation of hospital patient flow, the operations of medical tourism, and capacity and access decisions. Each leaves part of the gap this study addresses, and a final subsection states the gap they leave together.

\subsection{Hybrid simulation of patient flow}

Discrete-event simulation (DES) is widely used to model hospital patient flow, queueing, and resource use. \citet{Jacobson2013} survey its use across emergency departments, outpatient clinics, and inpatient units, and \citet{Vazquez2021} update that survey. When patients' own decisions change the demand a resource sees, agent-based simulation (ABS) represents them as actors whose states and choices change over time: \citet{Chan2010} set out the tutorial case, and \citet{Alibrahim2018} model patient choice of provider within accountable care organizations.

Hybrid models that combine the two methods are increasingly used in healthcare operations research, as the systematic review by \citet{Kar2025HybridHealthcare} documents. Two hybrid studies sit closest to the present one in structure. \citet{Kittipittayakorn2016} integrate discrete-event and agent-based simulation to improve an orthopedic outpatient department, with the agent layer representing the movement and interaction of patients and staff. \citet{Viana2018} build a hybrid model of an overdue-pregnancy outpatient clinic in which the discrete-event layer carries the clinic process and the agent layer each patient's progression between visits. More recent work extends the approach to inpatient access and to human factors: \citet{Landa2025EmergencyBoarding} propose a hybrid framework for reducing emergency patient boarding, where the constraint is access to inpatient beds, and \citet{Alhaider2025SituationAwareness} combine DES and agent-based modeling to study distributed situation awareness among clinical staff. These studies show how hybrid simulation can represent behavioral or organizational mechanisms alongside patient flow, with the mechanisms that matter differing by setting. None of them, however, addresses a medical-tourism setting or a department in which international and local patients are served from the same beds, specialists, and slots.

\subsection{Operations of medical tourism}

Much of the medical-tourism literature examines the growth and commodification of the phenomenon, barriers and competition in destination markets, destination attractiveness, medical tourism as one form of cross-border care, and telework in the industry \citep{Connell2006,Connell2013,Heung2011,Fetscherin2016,Sag2025MedTourismCompetition,Ziemba2025HealthcareGlobalization,Yorulmaz2026Telework}. This stream explains why patients travel but says little about how a hospital should configure beds, specialists, or appointment rules once they arrive.

A smaller operations-focused stream casts the destination medical center as a scheduling problem. \citet{Rezaeiahari2017} formulate an optimization model for scheduling traveling patients through a sequence of appointments, \citet{Rezaeiahari2020} extend it with a simulation-optimization approach that captures stochastic service times, and \citet{Bakhshandeh2024MedTourismScheduling} propose a multi-objective scheduling model for medical-tourism centers. These studies come closest to the present setting, but in each the traveling patient is modeled as a case to be scheduled, and behavior enters, if at all, as variability to schedule around; \citet{Zhou2022} make this explicit in an intraday scheduling model with patient re-entries and behavioral heterogeneity. In the operations models of medical tourism, patient behavior is therefore absent or imposed from outside rather than generated within the system.

\subsection{Capacity and access decisions}

On the inpatient side, bed-management research asks how much flexibility a hospital should allow in the use of its beds. In large systems, sharing a small number of beds captures most of the benefit of complete pooling \citep{Bekker2017FlexibleBeds}; a clustered overflow configuration, in which a patient turned away by a full primary ward is admitted to an overflow ward serving a designated group of specialties, compares well with dedicated and fully flexible wards when patients' willingness to wait is relatively short \citep{Izady2021ClusteredOverflow}, and bed allocation can be optimized within it \citep{Gong2022ClusteredOverflow}. These studies model a general inpatient population analytically or by optimization, and their common lesson is that how existing beds are made available matters in addition to how many there are. Access channels raise a similar question: \citet{KammrathBetancor2025OnlineBooking} find that the effect of online appointment scheduling on care efficiency depends on the context in which it is introduced. Each of these studies examines one lever at a time. Designed experiments vary several levers together, but they are rarely combined with DES: a recent review finds 86 such studies, seven of them in healthcare and more than half with full factorial designs \citep{Gjerloev2026DoEDES}. In healthcare, designed experiments have analyzed an outpatient orthopedic clinic \citep{Baril2014OrthoDOE} and the resource configuration of outpatient chemotherapy departments \citep{Corsini2024ChemoDOE}, and \citet{Keshtkar2020Boarding} pair one with a hybrid system-dynamics and DES model of an emergency department and acute medical unit to assess interventions against patient boarding. None of these experiments concerns a department that serves local and international patients from the same beds, specialists, and slots; in the reviewed literature, the capacity, channel, and priority levers of such a department have not been screened jointly.

\subsection{Research gap}\label{subsec:gap}

Table~\ref{tab:litmap} compares the closest studies on four features. None has more than one of them in full, and none serves medical tourists and local patients from the same resources. What is missing is the combination: a department shared by the two groups, patient behavior generated within the model, and a joint screening of the management decisions in a designed experiment. This study provides all three.

\begin{table*}[ht]
\centering
\caption{Features of the most closely related studies compared with the present work.}
\label{tab:litmap}
\centering
\footnotesize
\setlength{\tabcolsep}{2.5pt}
\renewcommand{\arraystretch}{1.15}
\begin{tabular}{@{}>{\raggedright\arraybackslash}p{5.7cm}>{\raggedright\arraybackslash}p{2.3cm}cccc@{}}
\toprule
Study & Method & \begin{tabular}[c]{@{}c@{}}Tourists\\and locals\end{tabular} & \begin{tabular}[c]{@{}c@{}}Patient\\behavior\end{tabular} & \begin{tabular}[c]{@{}c@{}}Shared\\beds\end{tabular} & \begin{tabular}[c]{@{}c@{}}Designed\\experiment\end{tabular} \\
\midrule
\citet{Kittipittayakorn2016} & Hybrid ABS+DES  &  &  &  &  \\
\citet{Viana2018} & Hybrid ABS+DES  &  & $\circ$ &  &  \\
\citet{Landa2025EmergencyBoarding} & Hybrid ABS+DES  &  &  & $\circ$ &  \\
\citet{Alhaider2025SituationAwareness} & Hybrid ABS+DES  &  &  &  &  \\
\citet{Keshtkar2020Boarding} & Hybrid SD+DES  &  &  & $\circ$ & $\checkmark$ \\
\citet{Alibrahim2018} & ABS  &  & $\checkmark$ &  &  \\
\citet{Bekker2017FlexibleBeds} & Analytical  &  &  & $\checkmark$ &  \\
\citet{Izady2021ClusteredOverflow} & Optimization  &  & $\circ$ & $\checkmark$ &  \\
\citet{Gong2022ClusteredOverflow} & Optimization  &  &  & $\checkmark$ &  \\
\citet{Baril2014OrthoDOE} & DES  &  &  &  & $\checkmark$ \\
\citet{Corsini2024ChemoDOE} & DES  &  &  &  & $\checkmark$ \\
\citet{Rezaeiahari2017} & Optimization  & $\circ$ &  &  &  \\
\citet{Rezaeiahari2020} & Sim.-optim.  & $\circ$ &  &  &  \\
\citet{Bakhshandeh2024MedTourismScheduling} & Optimization  & $\circ$ &  &  &  \\
\citet{Zhou2022} & Optimization  &  & $\circ$ &  &  \\
\midrule
This study & Hybrid ABS+DES  & $\checkmark$ & $\checkmark$ & $\circ$ & $\checkmark$ \\
\bottomrule
\end{tabular}
\vspace{2pt}

\parbox{0.96\linewidth}{\footnotesize Notes. ABS, agent-based simulation; DES, discrete-event simulation; SD, system dynamics. Tourists and locals: medical tourists and local patients served from the same resources; $\circ$ marks a medical-tourism setting that models the traveling patients only. Patient behavior: patient decisions (adherence, abandonment, provider switching, channel choice) endogenous to the flow; $\circ$ marks behavior represented as exogenous variability or as a single progression state. Shared beds: beds lent between wards; $\circ$ marks inpatient bed access modeled without inter-ward sharing. Designed experiment: management decisions varied jointly in an experimental design.}
\end{table*}

\section{Problem Description}\label{sec:problem}

This section states the problem in general terms: the shared service system, the patient behavior that shapes its demand, the management decisions and performance measures, and the scope and assumptions of the model.

\subsection{Shared service system}\label{subsec:system}

Consider a hospital department that offers $S$ clinical specialties, each run as a section with its own inpatient beds and specialist physicians; section $s$ has $b_s$ beds and $d_s$ specialists. Patients arrive at an average rate $\lambda$ per day and belong to one of two classes: local patients, and medical tourists, who make up a share $\rho$ of arrivals. Both classes may need the same specialists, appointment slots, and beds, and they differ mainly in the consequences of waiting: a longer queue delays a local patient's care, whereas for a tourist it also extends accommodation and travel and raises the chance that the planned treatment is abandoned.

\begin{figure*}[t]
\centering
\includegraphics[width=\textwidth]{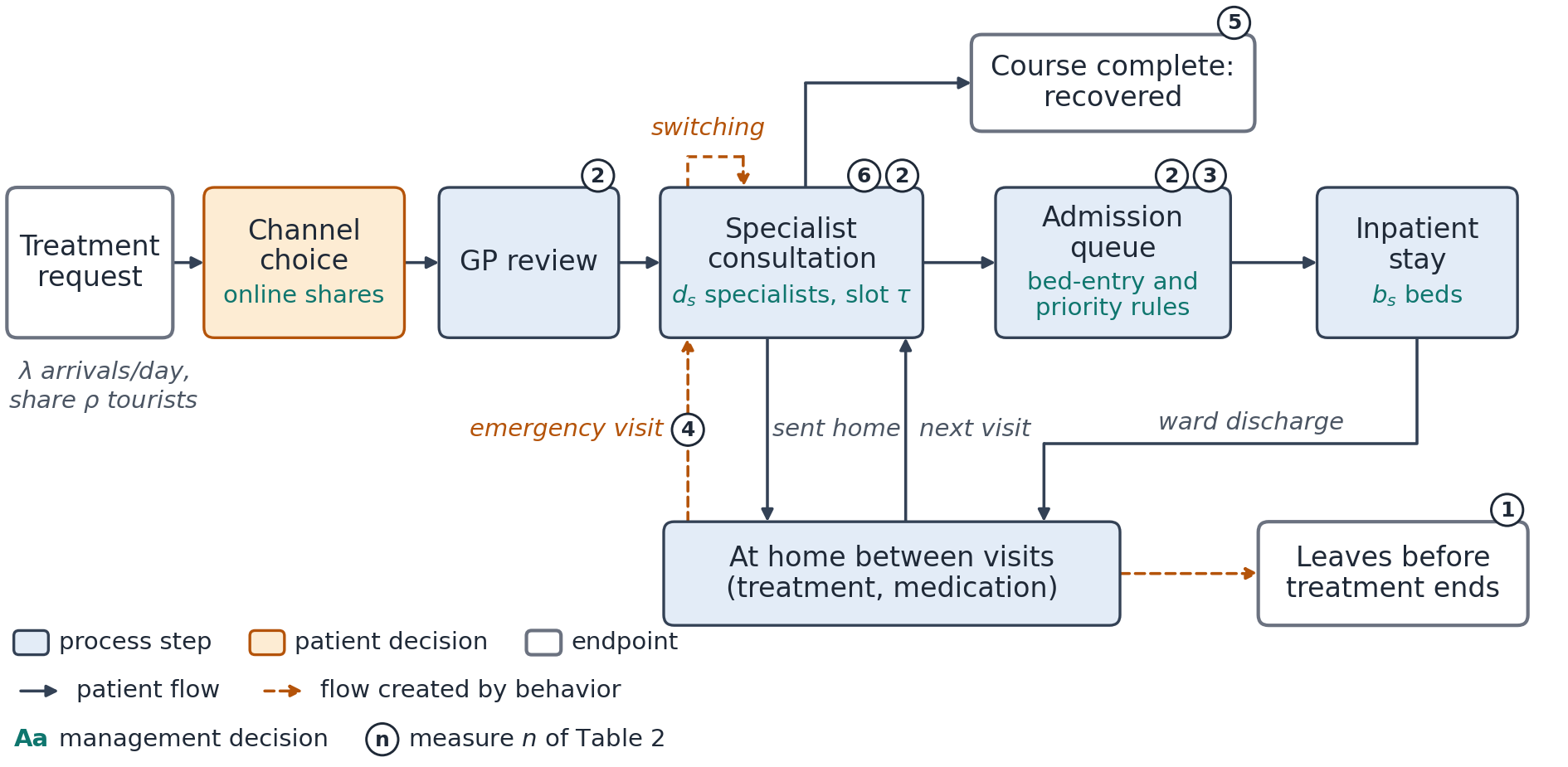}
\caption{Patient pathway through the shared department. Teal text names the management decision that acts at each step. Numbered circles mark where the measures of Table~\ref{tab:responses} are recorded: 1 early abandonment, 2 average wait in system (all queues), 3 medical-tourist admission wait, 4 emergencies before appointment, 5 recovered patients, 6 specialist utilization. Orange dashed arrows are flows created by patient behavior.}
\label{fig:pathway}
\end{figure*}

A patient enters through one of two channels, online or in person. A pool of $n_{\mathrm{GP}}$ general practitioners, available continuously, reviews each request and refers the patient to the matching specialist section. Specialists follow fixed weekly patterns of clinic hours, hospital rounds, and file-review windows and see patients in appointment slots of length $\tau$, reserved separately for online and in-person visits. A consultation ends in discharge, treatment at home, or a request for admission; an admitted patient waits for a bed in the relevant section, occupies it for the length of stay, and is discharged. Most conditions need a course of several consultations, so a patient returns for later visits until the course is complete. Bed capacity is limited and assigned by section, so the difficulty lies both in how much capacity the department has and in how it is made available across patients, sections, and channels. Figure~\ref{fig:pathway} traces this pathway and marks where the management decisions act and where the performance measures of Table~\ref{tab:responses} are recorded.

\subsection{Patient behavior}\label{subsec:behavior}

Patient behavior makes this system more than a sequence of queues. A patient chooses the entry channel according to preference and circumstance, chooses a specialist on reputation and queue conditions and may later switch, and may abandon the system after a poor experience such as a long wait or a contested recommendation. A patient under medication moves between adherent and non-adherent states, and a lapse combined with worsening health triggers an unscheduled emergency consultation before the planned visit. Each behavior changes the demand the department sees: channel choice loads a different registration queue, switching moves demand between specialist queues, abandonment removes demand, and emergency re-entry adds it.

\subsection{Management decisions and performance measures}\label{subsec:decisions}

Management controls four groups of decisions: capacity, the number of beds and of specialists in each section; channel design, the shares of tourists and of local patients routed to online consultation; scheduling and priority rules, how patients from each channel are scheduled into beds and which class receives priority in the admission queue; and appointment timing, the length of a clinic slot. The decisions interact, since stronger priority for tourists may lengthen the local wait and moving patients online may shift workload onto that channel.

Arrivals, service times, lengths of stay, and every behavioral transition are random, the resources are shared across sections and classes, and the behaviors feed back into the queues they load, so performance is evaluated by simulation. Table~\ref{tab:responses} defines the six performance measures, each a run-level statistic collected after a warm-up period. Because the measures can favor different settings, the aim is to identify the decisions with the strongest effects and the trade-offs among them, not to optimize any one measure.

\begin{table*}[ht]
\centering
\caption{Performance measures.}
\label{tab:responses}
\centering
\small
\begin{tabular}{@{}>{\raggedright\arraybackslash}p{3.7cm}>{\raggedright\arraybackslash}p{8.6cm}p{1.7cm}@{}}
\toprule
Measure & Operational meaning & Direction \\
\midrule
Early abandonment & Number of patients leaving the system before completing treatment because of dissatisfaction or excessive waiting & Minimize \\
Average wait in system & Mean waiting time across all patients and all queues in the system & Minimize \\
Medical-tourist admission waiting time & Mean time spent by medical tourists waiting for hospital admission & Minimize \\
Emergencies before appointment & Number of patients whose condition deteriorates before scheduled service & Minimize \\
Recovered patients & Number of patients discharged in recovered condition over the simulation reporting horizon & Maximize \\
Specialist utilization & Utilization percentage of specialist physicians, reported by specialty & Maximize \\
\bottomrule
\end{tabular}
\end{table*}

\begin{figure*}[t]
\centering
\includegraphics[width=\textwidth]{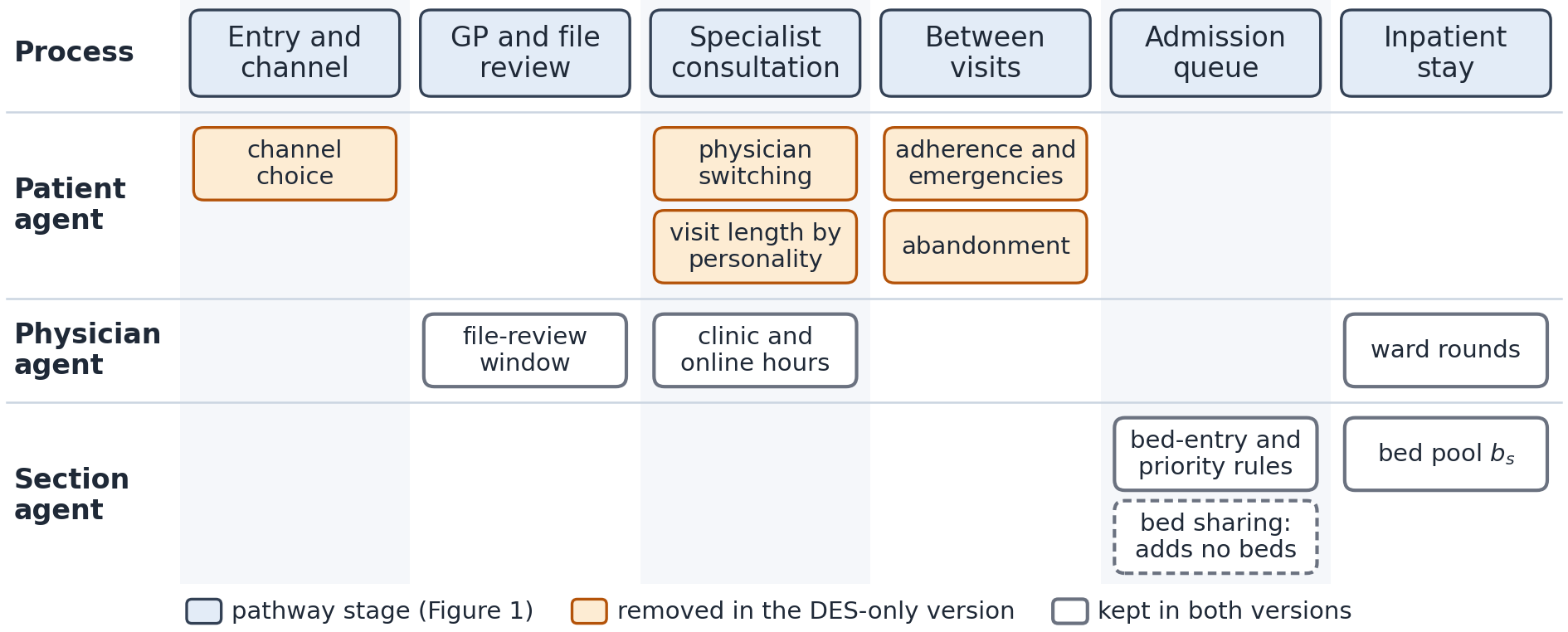}
\caption{Mechanisms of the agent layer by stage of the patient pathway (Figure~\ref{fig:pathway}). Orange mechanisms are the patient behaviors that the DES-only version removes; grey mechanisms are present in both versions.}
\label{fig:arch}
\end{figure*}

\subsection{Scope and assumptions}

The model covers the parts of the department that bear directly on the decisions studied; Table~\ref{tab:assumptions} lists its assumptions, including what lies outside it.

\begin{table*}[ht]
\centering
\caption{Modeling assumptions.}
\label{tab:assumptions}
\begin{tabular}{@{}p{4.3cm}p{11.2cm}@{}}
\toprule
Item & Assumption \\
\midrule
Service scope & $S$ specialties, each a section with its own beds and specialists; nurses, imaging, and inter-hospital referral are outside the model \\
Arrivals & Random arrivals at mean rate $\lambda$ per day; a share $\rho$ are medical tourists \\
Initial triage & $n_{\mathrm{GP}}$ general practitioners available continuously for review and routing \\
Specialist availability & Fixed weekly schedules combining clinic hours, hospital rounds, and file review; appointment slots of length $\tau$, reserved separately for online and in-person visits \\
Service times & Triangular distributions by specialty and visit type; duration also depends on patient attributes \\
Treatment course & Each patient requires a course of consultations whose length is drawn from a condition-specific distribution; a patient counts as recovered only after completing the course \\
Patient attributes & Age, gender, disease type, personality trait, online-visit preference, and hospitalization preference affect behavior \\
Physician choice & Patients choose among available physicians on reputation and queue conditions and may later switch \\
Emergency escalation & Poor medication adherence and worsening health status can trigger an unscheduled emergency consultation \\
Medication-adherence transitions & Transition rules between adherence and health states specified from field-study inputs, expert interviews, and a questionnaire-based elicitation \\
Bed sharing & Limited to designated compatible sections \\
Capacity within a run & Bed counts $b_s$ and specialist counts $d_s$ fixed for the duration of a run \\
Priority & Admission priority between local patients and medical tourists is set by the priority-policy lever; tourists come first in the released model \\
Horizon & Fixed run length with an initial warm-up excluded from all statistics \\
\bottomrule
\end{tabular}
\end{table*}

\section{Methodology}\label{sec:method}

This section presents the hybrid simulation model and the evidence used to check it, the screening experiment over the management factors, the comparison with a DES-only version of the model, and the use of large language models in the study.

\subsection{Hybrid simulation model}\label{subsec:framework}

The model is implemented in AnyLogic and combines DES with ABS \citep{Borshchev2013AnyLogic}. DES represents the sequence of care: registration, triage, consultation, admission, inpatient treatment, and discharge. ABS represents patients, physicians, and hospital sections as agents whose decisions and states change how patients move through that sequence. This paper calls that part the agent layer.

Figure~\ref{fig:arch} shows where each agent acts on the process and which of its mechanisms the DES-only version removes. The process layer follows the pathway of Figure~\ref{fig:pathway}. Online Resource 1 documents the implementation: the flowcharts, the agent state-charts with their parameter tables, including the exact bed-transfer rule, and the service times and weekly specialist schedules.

\subsubsection{Patient behavior}

Each patient is an agent whose attributes (class, disease type, age, gender, personality trait, and preferences for online consultation and hospitalization) drive the patient behaviors, including the acceptance or rejection of a recommendation; its medication-adherence and health states generate the unscheduled consultations, which re-enter the process layer ahead of the planned appointment. Patient characteristics also lengthen consultations: an anxious or contesting patient occupies the slot for longer.

\subsubsection{Physician availability}

Physician agents carry the specialists' weekly schedules, moving among home, clinic, hospital-round, online-visit, in-person-visit, and file-review states; the process layer books a consultation only when the agent is in a state that allows that type of service. A popularity mechanism, updated from patients' experience and queue conditions, shapes which physician a patient turns to next.

\subsubsection{Hospital sections and bed sharing}\label{subsec:beds}

Each hospital-section agent holds its bed pool and admission queue, tracks occupancy, applies the admission and priority rules, and answers admission requests from the process layer. Waiting patients are moved into free beds only when a new patient joins the section's admission queue, not when a patient is discharged, so a freed bed can stay empty until the section's next admission request. The released model also contains a daily routine meant to let a congested section in the compatible group borrow idle beds from another. As implemented, the routine never adds beds: its transfer is zero or negative, and a negative transfer lowers the borrowing section's bed count for the rest of the run. Both rules are the same in the hybrid and DES-only models, and the run logs do not record section bed counts, so how often the routine lowered capacity is unknown; the results in this paper do not evaluate bed sharing.

\subsection{Model checking and case-based evidence}\label{subsec:vv}

Following standard simulation guidance \citep{Sargent2010,RobinsonSimulationBook} and the reporting practice recommended for healthcare simulation \citep{Monks2018STRESS,Zhang2020ReportingQuality}, three kinds of evidence are reported separately: whether the model's structure matches the care process, whether its inputs are grounded in the case setting, and whether its simulated behavior is operationally reasonable.

The conceptual structure comes from interviews with physicians and staff at the case-study hospital and at a separate clinic, who defined and reviewed the patient pathway, the access channels, triage, specialist schedules, inpatient admission, priority rules, compatible-section bed sharing, and the pathway through which deterioration generates an unplanned consultation. The medication-adherence component is informed by the field study, expert input, and questionnaire-based elicitation; the number and roles of interviewees and respondents were not recorded in a reportable form, which limits the reproducibility of the elicitation. The model inputs are tied to the same setting. Finally, queue levels and throughput are examined over time to confirm that the system settles into a stable operating pattern.

This evidence supports face validity and case grounding rather than a claim of predictive validation against an independent hospital dataset.

\subsection{Screening the management factors}\label{subsec:doe}

A two-level fractional factorial design identifies which management factors have the strongest effects on the six performance measures within the hybrid model; its design, factors and levels, screening model, and joint reading across measures are described in turn.

\subsubsection{Experimental design}

Sixteen factors give $2^{16}=65{,}536$ combinations in a full two-level factorial experiment; the study uses a 256-run fraction, $1/256$ of the full design, generated with the \texttt{FrF2} package \citep{Groemping2014}. The design has resolution~V: estimated main effects are not mixed with two-factor interactions, and two-factor interactions are not mixed with one another, though both can be mixed with higher-order interactions, which are assumed small under the usual effect-sparsity principle \citep{BoxHunterHunter2005,Montgomery2017DOE}. The shortest word in the design's defining relation has length five, which is what resolution~V requires.

\subsubsection{Factors and levels}

The 16 factors, A--P, cover the four decision groups. Factors A--E are the bed counts $b_s$ by section and F--J the specialist counts $d_s$; K and L are the shares of tourist and local patients routed to online consultation; M and N are the admission-scheduling rules for the online and in-person channels, and O is the patient-priority rule; and P is the clinic appointment-slot interval $\tau$. Table~\ref{tab:factors} reports the low and high levels and the case-study baseline values; the results apply only within the tested ranges.

\begin{table*}[ht]
\centering
\caption{Factor definitions, candidate low/high levels, and baseline operating values.}
\label{tab:factors}
\centering
\footnotesize
\setlength{\tabcolsep}{4pt}
\renewcommand{\arraystretch}{1.15}
\begin{tabular}{@{}cp{4.4cm}p{1.3cm}p{1.3cm}p{1.1cm}p{1.6cm}@{}}
\toprule
Factor & Definition & Low & High & Baseline & Units / coding \\
\midrule
A & Number of beds in section 1 & 32  & 52   & 40 & Beds \\
B & Number of beds in section 2 & 40  & 65   & 50 & Beds \\
C & Number of beds in section 3 & 56  & 91   & 70 & Beds \\
D & Number of beds in section 4 & 16  & 26   & 20 & Beds \\
E & Number of beds in section 5 & 48  & 78   & 60 & Beds \\
F & Specialists in section 1   & 1   & 2    & 1  & Specialists \\
G & Specialists in section 2   & 4   & 8    & 6  & Specialists \\
H & Specialists in section 3   & 1   & 3    & 2  & Specialists \\
I & Specialists in section 4   & 1   & 3    & 2  & Specialists \\
J & Specialists in section 5   & 2   & 4    & 3  & Specialists \\
K & Share of tourist patients on online consultation & 20\% & 60\% & n/a & \% of tourists \\
L & Share of local patients on online consultation & 10\% & 40\% & n/a & \% of locals \\
M & Online bed-scheduling rule & Direct & Deferred & Deferred & Categorical $\pm 1$ \\
N & In-person bed-scheduling rule & Direct & Deferred & n/a & Categorical $\pm 1$ \\
O & Admission-queue priority rule & Tourists first & Referral order & Tourists first & Categorical $\pm 1$ \\
P & Clinic appointment-slot interval & 2~min & 5~min & 5~min & Min.\ per slot \\
\bottomrule
\end{tabular}
\vspace{2pt}

\parbox{0.96\linewidth}{\footnotesize Notes. Baseline values are the case-study operating values; for $M$ and $O$, the rule in the released model. $M$ and $N$ set how online and in-person referrals enter a section's bed list: Direct at the next free position, Deferred by a fixed number of days when the bed queue exceeds a threshold. $O$ low puts medical tourists first and online referrals last; $O$ high admits in referral order. The factor-to-model mapping is documented with the code.}
\end{table*}

\subsubsection{Screening model}

For each performance measure, the simulation outputs are analyzed with a linear screening model in the main effects and retained two-factor interactions, following the standard $2^k$ fractional factorial screening framework \citep{BoxHunterHunter2005,Montgomery2017DOE}:
\begin{equation}
Y = \beta_0 + \sum_{i=1}^{16} \beta_i x_i + \sum_{i<j}\beta_{ij}x_ix_j + \varepsilon,
\label{eq:rsm}
\end{equation}
where $x_i$ is the coded level of factor $i$, $\beta_i$ its main effect, and $\beta_{ij}$ the interaction between factors $i$ and $j$. Because the resolution-V design estimates every two-factor interaction clear of the main effects, terms are retained on their own significance: a main effect or two-factor interaction is kept when its coefficient is significant at the 5\% level, no hierarchy constraint is imposed, three-factor and higher interactions are excluded, and no adjustment for multiple testing is applied. Each response is screened over 16 main effects and 120 two-factor interactions, so about forty of the 816 terms across the six responses are expected to be retained by chance alone; a small retained term is a candidate for confirmation, not an established effect. The design is unreplicated, one run per design point, so the residual term contains both simulation variability and any higher-order effects not in the fitted model; the results are read for the direction and relative importance of effects, with magnitudes treated as approximate.

\subsubsection{Reading the factors across measures}

The six measures do not always favor the same setting, so the screening is read jointly under the case study's priority order, assigned before the final settings are interpreted. When the main-effect directions agree across the higher-priority measures, the choice is straightforward; when they conflict, the relevant two-factor interactions and fitted surfaces are examined before a case-specific level is selected.

\subsection{Hybrid versus DES-only comparison}\label{subsec:protocol}

The hybrid model is compared with a DES-only version under the same fixed operating configuration. The DES-only model keeps the patient-flow structure, the resource configuration, and the bed-sharing routine, but removes the behavioral mechanisms of the agent layer: abandonment, physician switching, behavior-dependent channel choice, medication-adherence effects, and behavior-dependent consultation duration. Every other parameter stays in place.

Because these mechanisms are removed together, the comparison measures their combined effect at that operating point, not the effect of any one of them. The bed-sharing routine responds to the length of a section's bed queue, which differs between the two models, so it may have lowered capacity by different amounts in each; the run logs can neither confirm nor rule this out.

The two variants are run independently under randomized seeds, and means with 95\% confidence intervals are reported across replications. Differences are tested with Welch's two-sample $t$-test, which does not require equal variances or replication counts; the count-valued measures are treated as run-level aggregates, for which the central-limit approximation is reasonable; Mann--Whitney tests are reported as a distribution-free check; and Holm's procedure adjusts for multiple comparisons. The inpatient admission waiting time is the prespecified primary outcome, because the comparison configuration is chosen to make inpatient beds the bottleneck.

\subsection{Use of large language models}\label{subsec:llm}

Large language models (Claude, Anthropic; ChatGPT, OpenAI) were used under the authors' direction to restructure and edit the manuscript text, to write the figure and analysis scripts, to recompute the reported comparison statistics from the simulation run logs, to assist with scripting and execution of the robustness runs under the authors' supervision, to inspect the simulation code against its documentation, and to check the reference list against the Crossref database. They were not used to build the simulation model or to determine the conclusions. Every reported statistic was independently checked against the original simulation logs. The authors reviewed every AI-assisted passage and computation and take full responsibility for the content.

\section{Numerical Results}\label{sec:results}

This section describes the case study, reports the main effects and interactions of the management factors and the settings they suggest, and compares the hybrid and DES-only models at a bed-constrained configuration.

\subsection{Case study and experimental setting}\label{subsec:case}

\begin{table*}[ht]
\centering
\caption{Case-study parameters.}
\label{tab:caseparams}
\centering
\footnotesize
\setlength{\tabcolsep}{5pt}
\renewcommand{\arraystretch}{1.15}
\begin{tabular}{@{}p{4.3cm}p{7.3cm}@{}}
\toprule
Parameter & Value \\
\midrule
Specialties ($S=5$), in section order & 1 cardiology, 2 internal medicine, 3 cosmetic surgery, 4 pediatrics, 5 breast oncology \\
Arrival rate $\lambda$ & 24 patients per day (screening and robustness runs); 16 (model comparison) \\
Tourist share $\rho$ & About one tourist per ten local patients ($\rho\approx 1/11$) \\
General practitioners $n_{\mathrm{GP}}$ & 3, continuously available \\
Beds by section $b_s$ & 40, 50, 70, 20, 60 (screening baseline); 30, 16, 22, 2, 16 (model comparison) \\
Specialists by section $d_s$ & 1, 6, 2, 2, 3 (screening baseline); 3, 5, 3, 3, 5 (model comparison) \\
Clinic slot interval $\tau$ & 5 minutes (screening baseline); 2 minutes (low level of $P$; model comparison) \\
Compatible bed-sharing group & Cardiology, internal medicine, breast oncology; cosmetic surgery and pediatrics do not share \\
Service times and schedules & Triangular service times by specialty and visit type; fixed weekly specialist schedules \\
Horizon & 365 days per screening run, 300 per comparison run; 10-day warm-up \\
\bottomrule
\end{tabular}
\end{table*}

The case is an international patient department in one hospital in Iran, developed from information collected from physicians and staff at Takhte Jamshid Hospital and at Mehr Clinic, a separate organization. The department covers $S=5$ specialties, numbered as sections 1 to 5 throughout the paper: cardiology, internal medicine, cosmetic surgery, pediatrics, and breast oncology. Table~\ref{tab:caseparams} gives the parameters, including the screening baseline documented in the case-study source (the baseline column of Table~\ref{tab:factors}) and the bed-constrained configuration used for the model comparison. The department allows bed sharing only among clinically compatible sections: cardiology, internal medicine, and breast oncology may share beds, while cosmetic surgery, whose workload is elective, and pediatrics, whose patients need a dedicated ward, keep their beds. The screening and the comparison run at different arrival rates and bed counts because they answer different questions, so their absolute values are not comparable across stages.

The six performance measures are weighed in the following order: (i)~minimizing the medical-tourist admission waiting time, (ii)~minimizing the average wait in the system, (iii)~preventing early abandonment, then (iv)~reducing emergencies before appointment, (v)~increasing recovered patients, and (vi)~increasing specialist utilization. Tourist waiting time leads the order because a tourist's delay carries accommodation and travel costs on top of the cost of delayed care. Specialist utilization, last in the order, is also read in the model comparison as a diagnostic of spare consultation capacity.

\subsection{Main effects of the management factors}\label{subsec:screening}

\begin{figure*}[ht]
\centering
\includegraphics[width=\textwidth]{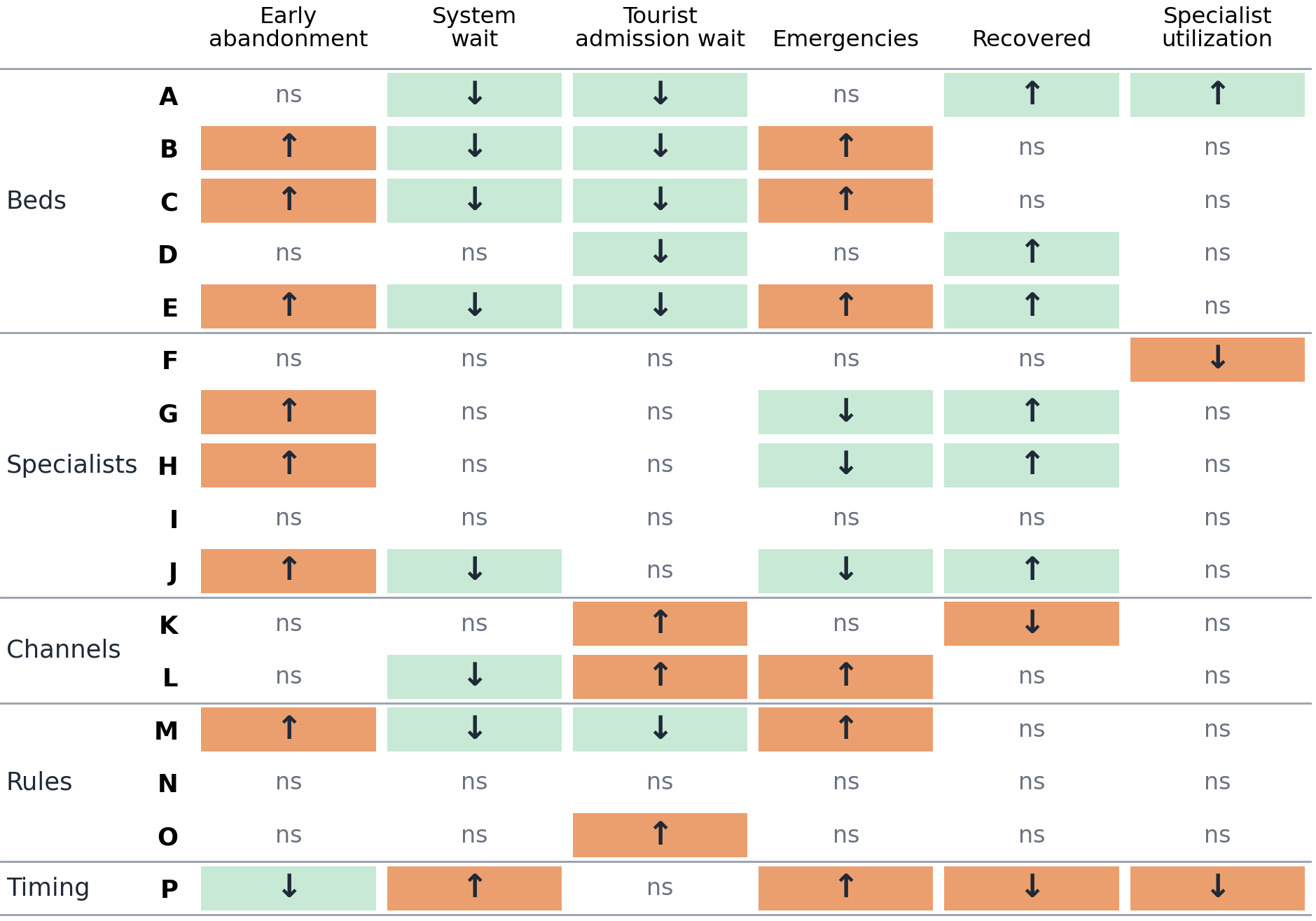}
\caption{Direction of the retained main effects, grouped by decision type. An arrow gives the change in the measure when the factor moves from its low to its high coded level; green cells improve the measure, orange cells worsen it, and \textit{ns} marks no retained effect.}
\label{fig:matrix}
\end{figure*}

Figure~\ref{fig:matrix} summarizes the direction of each retained main effect, grouped by decision type.

The clearest pattern is in inpatient capacity: raising bed capacity reduces the medical-tourist admission waiting time in every section and the system wait in four of the five, which makes bed capacity the most consistent driver of queue performance. In internal medicine, cosmetic surgery, and breast oncology, higher bed levels also raise early abandonment and emergencies. Specialist capacity acts elsewhere: raising specialist counts has little effect on the two waiting measures (only $J$ lowers the system wait) and works instead through emergencies, recovered patients, abandonment, and utilization. The channel and policy factors enter through both main effects and interactions with signs that conflict across measures: a higher tourist online share ($K$) lengthens the tourist queue and lowers recovered patients, because online-referred patients rank behind in-person referrals in the admission queue; a higher local online share ($L$) shortens the system wait but lengthens the tourist queue and raises emergencies; the online bed-scheduling rule ($M$) shortens both waits while raising abandonment and emergencies; and the priority rule ($O$) acts on the tourist queue mainly through interactions. The interaction surfaces resolve these conflicts. Finally, the slot interval ($P$) carries a direct trade-off: at its 2-minute level it lowers the system wait and emergencies and raises recovered patients and utilization, at the cost of higher early abandonment.

The residuals are well behaved for the abandonment, system-wait, and tourist-admission-wait models and heavier-tailed for the other three (plots in Online Resource 1).

\subsection{Interactions and case-study settings}\label{subsec:rsm}

Because each factor has two tested levels, the interaction surfaces (Online Resource 1) show the direction of a fitted interaction, not curvature, and their extreme values fall at combinations of the tested levels.

For medical-tourist admission waiting time, the most informative interactions pair the patient-priority rule with the online bed-scheduling rule ($MO$) and with the shares of tourist and local patients on online consultation ($KO$ and $LO$). In the $MO$ surface the measure is lowest when $M$ is high and $O$ is low, and the $KO$ and $LO$ surfaces are both lowest when the online share and $O$ are at their low levels; all three match the per-factor signs. Under the tested admission rules, online access lengthens the tourist admission wait, and its benefit to the system wait comes from the local channel.

Two further groups of interactions decide the remaining conflicts. For emergencies, specialist capacity interacts with the slot interval ($FP$, $GP$, $JP$): the count is lowest when $P$ sits at its low level, after which the specialist count carries a smaller marginal effect, so tightening the slot interval is the dominant lever within the tested range. For specialist utilization, bed and specialist capacity interact ($AF$): utilization is highest when the bed count is high and the specialist count low, which shows that added specialist capacity yields little when inpatient beds restrict flow.

\subsubsection{Factor settings under the priority order}\label{subsec:doe_rules}

\begin{table}[ht]
\centering
\caption{Case-study setting of each factor under the priority order: ``$+$'' high, ``$-$'' low, ``$0$'' no retained effect. Levels are defined in Table~\ref{tab:factors}; directions are those of Figure~\ref{fig:matrix}.}
\label{tab:final}
\centering
\footnotesize
\setlength{\tabcolsep}{2.5pt}
\renewcommand{\arraystretch}{1.15}
\begin{tabular}{@{}lcccccccccccccccc@{}}
\toprule
Factor & A & B & C & D & E & F & G & H & I & J & K & L & M & N & O & P \\
\midrule
Level & $+$ & $+$ & $+$ & $+$ & $+$ & $-$ & $-$ & $-$ & $0$ & $+$ & $-$ & $-$ & $+$ & $-$ & $-$ & $-$ \\
\bottomrule
\end{tabular}
\end{table}

Table~\ref{tab:final} sets each factor by the case study's priority order. The bed counts $A$--$E$ are set high because they shorten the top-priority medical-tourist admission wait. Where measures conflict, the higher-priority one decides: $G$ and $H$ are set low because they raise early abandonment, $J$ high because it shortens the system wait, and $L$ low because the tourist wait outranks the system wait. $F$ is set low because it lowers utilization, $I$ has no retained effect, and $K$, $M$, $N$, $O$, and $P$ follow their main effects and surfaces. The low level of $O$, the rule in the released model, keeps medical tourists ahead in the admission queue, an advantage that online referral removes. These are illustrative settings within the tested ranges, not optimal policies.

\begin{table*}[ht]
\centering
\caption{Hybrid versus DES-only model at the comparison configuration: means with 95\% confidence intervals over 300-day runs ($n=30$ hybrid, $n=36$ DES-only); differences tested with Welch $t$-tests.}
\label{tab:comparison}
\centering
\small
\setlength{\tabcolsep}{4pt}
\renewcommand{\arraystretch}{1.25}
\begin{tabular}{@{}>{\raggedright\arraybackslash}p{4.6cm}ccc@{}}
\toprule
Measure & Hybrid & DES-only & Difference \\
 & (mean, 95\% CI) & (mean, 95\% CI) & (95\% CI) \\
\midrule
Inpatient admission waiting time (days) & 14.8 (14.0, 15.7) & 25.8 (24.6, 27.0) & $-11.0$ $(-12.4, -9.5)^{***}$ \\
Medical-tourist admission waiting time (days)$^{\ddagger}$ & 8.1 & 4.7 & not tested$^{\ddagger}$ \\
Average wait in system (days) & 0.59 (0.56, 0.62) & 0.89 (0.84, 0.94) & $-0.30$ $(-0.36, -0.24)^{***}$ \\
Early abandonment (patients/run) & 120 (116, 123) & 0 & $+120$ (structural)$^{\dagger}$ \\
Emergencies before appointment (count/run) & 481 (442, 520) & 195 (176, 214) & $+286$ $(+243, +329)^{***}$ \\
Recovered patients (count/run) & 2{,}155 (2{,}110, 2{,}200) & 3{,}243 (3{,}199, 3{,}287) & $-1{,}088$ $(-1{,}149, -1{,}026)^{***}$ \\
Mean specialist utilization (\%) & 20.5 (19.9, 21.1) & 15.9 (15.8, 16.0) & $+4.7$ $(+4.1, +5.2)^{***}$ \\
\bottomrule
\end{tabular}
\vspace{2pt}

\parbox{0.96\linewidth}{\footnotesize Notes. Difference is the hybrid mean minus the DES-only mean, from unrounded means, so it can differ from the cell entries by one unit in the last digit. $^{***}p<0.001$. $^{\dagger}$~Structural: the DES-only model cannot generate abandonment. $^{\ddagger}$~Not tested: the DES-only model never generates the abandoning and re-presenting tourists who wait longest, so the two means do not draw on the same patients.}
\end{table*}

\begin{figure}[t]
\centering
\includegraphics[width=\columnwidth]{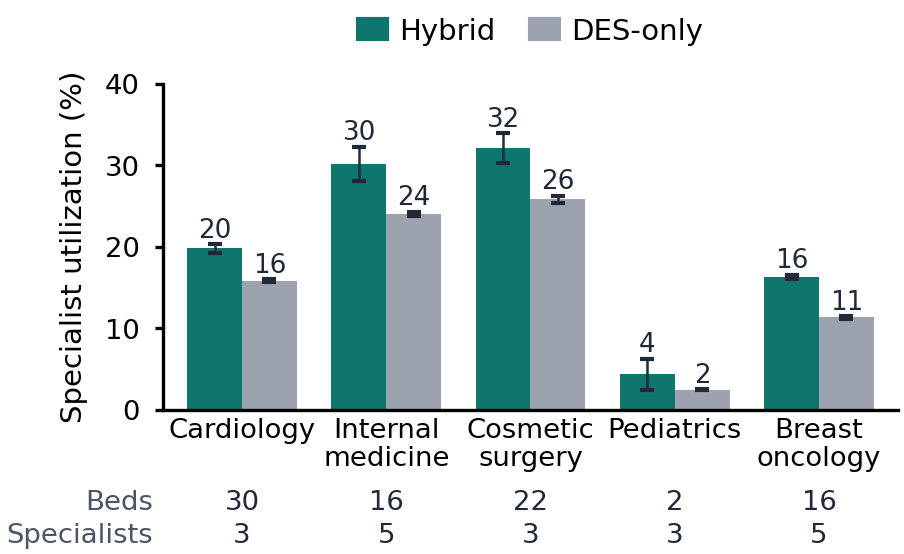}
\caption{Specialist utilization by specialty at the comparison configuration in the hybrid ($n=30$) and DES-only ($n=36$) models: means with 95\% confidence intervals. The rows below the axis give each specialty's beds and specialists in this configuration.}
\label{fig:util}
\end{figure}

\begin{figure*}[t]
\centering
\includegraphics[width=\textwidth]{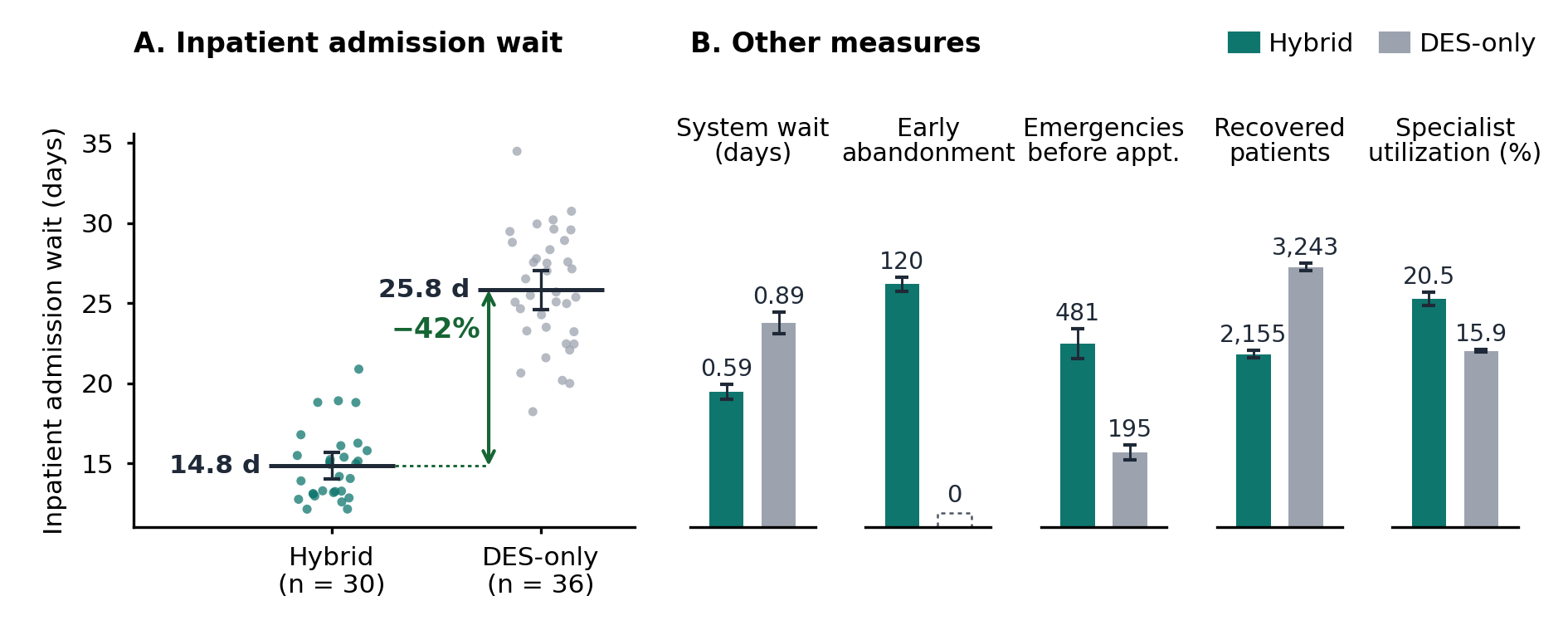}
\caption{Hybrid versus DES-only model at the comparison configuration. Panel A shows every replication of the inpatient admission wait with the mean and its 95\% confidence interval ($n=30$ hybrid, $n=36$ DES-only). Panel B shows the other measures as means with 95\% confidence intervals; the dashed outline marks a structural difference, since the DES-only model cannot generate abandonment, and is not a tested performance gain.}
\label{fig:results_main}
\end{figure*}

\subsection{Performance at the comparison configuration}\label{subsec:simoutputs}

The comparison configuration is a deliberately bed-constrained operating point, not the screening baseline. It sets the section bed counts to 30, 16, 22, 2, and 16 and the specialist counts to 3, 5, 3, 3, and 5, the values held in the database released with the model, with the clinic slot interval at 2 minutes and the arrival rate at 16 patients per day. All five bed counts lie below the low levels of the screening design (32, 40, 56, 16, and 48), so the configuration creates a bed-capacity bottleneck, while the specialist counts match or exceed the baseline in four of five sections. In the 30 hybrid runs, specialist utilization (realized consultation and ward-round time divided by scheduled hours) ranges from roughly 4\% to 32\% across the five sections (Figure~\ref{fig:util}), with a mean of about 20\%, despite a multi-week admission queue. The emergency count is a simulated event count from the adherence and deterioration pathway, not an empirical incidence estimate, and the recovered count covers completed treatment courses only, so it is not comparable with the roughly 4,640 arrivals over the reporting horizon.

The simulation also reaches a stable operating pattern over the reporting horizon: the admission queue settles instead of growing, online and in-person visits arrive in comparable daily volumes, and recovered-patient counts follow a recurring weekly cycle.

\subsection{Effect of removing the behavioral mechanisms}\label{subsec:comparison}

Table~\ref{tab:comparison} and Figure~\ref{fig:results_main} compare the two models, replicated 30 and 36 times at the same operating configuration; the unequal counts are incidental to the run schedule.

The primary outcome, inpatient admission waiting time, is 14.8~days (95\% CI 14.0 to 15.7) in the hybrid model against 25.8~days (24.6 to 27.0) in the DES-only model, a difference of 11.0~days (9.5 to 12.4) or about 42\%. Average waiting across the system also falls, from 0.89 to 0.59~days. Both differences are statistically significant ($p<0.001$).

The comparison removes several behavioral mechanisms at once, so the 42\% difference is their combined effect, with the caveat that the bed-sharing routine may have lowered capacity differently in the two models. Neither variant has been compared against observed waiting times in the department, so the difference measures sensitivity to a modeling choice; it is not an estimate of the waiting a hospital can save.

The hybrid model records about 120 abandonments per run and the DES-only model none, because abandonment is not represented there. Emergencies are higher in the hybrid model (481 versus 195) and recovered patients fewer (2{,}155 versus 3{,}243): the discrete-event variant keeps the deterioration mechanism, which still produces 195 emergencies, but not the agent layer's adherence dynamics, which generate the further escalations. The DES-only model's more favorable counts reflect these omitted pathways and should not be read as better system performance. Specialist utilization is likewise higher in the hybrid model, in every specialty (Figure~\ref{fig:util}; 20.5\% versus 15.9\% on average), since removing behavior-driven re-entry leaves the DES-only model with less specialist workload.

The tourist-specific admission wait, about 8.1~days in the hybrid model and 4.7~days in the DES-only model, is reported for completeness but not tested, because the two models do not contain the same contributing patient population (see the note to Table~\ref{tab:comparison}).

A Holm--Bonferroni adjustment across the five tested measures leaves every $p$-value below $0.001$, and two-sided Mann--Whitney tests agree (Cohen's $d=3.6$ for the inpatient waiting time). These intervals and $p$-values describe variation across simulation runs, not uncertainty about the real hospital.

\begin{figure}[t]
\centering
\includegraphics[width=\columnwidth]{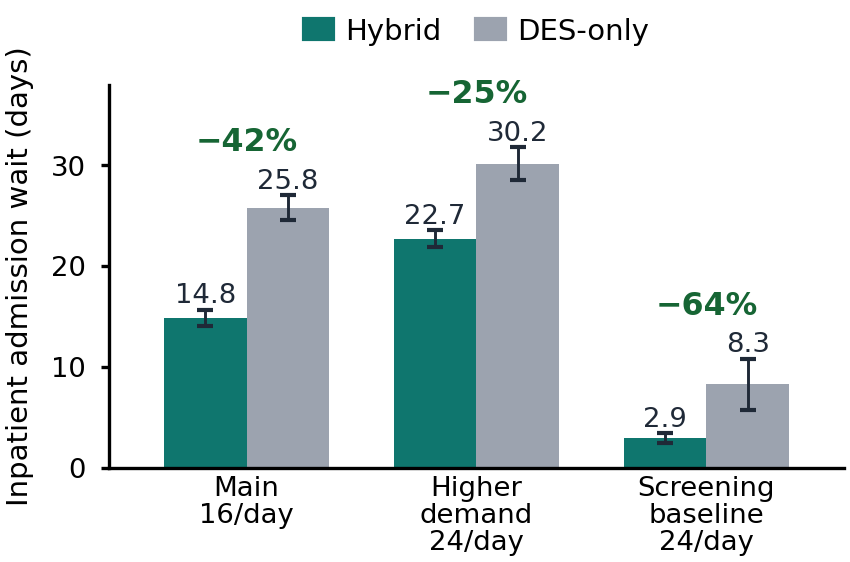}
\caption{Inpatient admission wait in the hybrid and DES-only models under the three comparison conditions: the bed-constrained configuration at 16 patients per day (main comparison; hybrid $n=30$, DES-only $n=36$) and at 24 patients per day, and the screening baseline's bed and specialist counts at 24 patients per day ($n=30$ each). Bars show means with 95\% confidence intervals; percentages give the hybrid model's difference.}
\label{fig:robust}
\end{figure}

\paragraph{Robustness across demand level and bed capacity.} The comparison is repeated with 30 runs of each model under two further conditions (Figure~\ref{fig:robust}; full results in Online Resource 1). At 24 patients per day on the bed-constrained configuration, the inpatient admission wait is 22.7 days (95\% CI 21.9 to 23.5) in the hybrid model against 30.2 days (28.5 to 31.8) in the DES-only model, about 25\% lower (Holm-adjusted $p<0.001$); the other measures differ as before, except the system wait ($p=0.067$). With the screening baseline's bed and specialist counts at 24 patients per day, both admission queues shrink, yet the wait is 2.9 days (2.5 to 3.4) against 8.3 days (5.8 to 10.8), about 64\% lower (adjusted $p<0.001$), with the DES-only queue far more variable across runs; the medical-tourist wait no longer differs clearly (2.6 versus 3.7 days). The inpatient-waiting difference therefore persists at both demand levels and on both sides of the bed constraint.

\section{Managerial Implications}\label{sec:managerial}

The numerical values are specific to the case study, but four implications may also apply to hospitals that serve local and international patients from shared resources: on inpatient capacity, on channel design and admission priority, on appointment timing, and on the structure of planning models.

\paragraph{Inpatient capacity and specialist expansion.} The strongest pattern in the case study is inpatient congestion beside modest specialist utilization. The same pattern appears in the screening, which ran at the larger baseline capacity: the bed counts drive both waiting measures, while the specialist counts have no retained effect on the medical-tourist admission wait. Adding physicians alone does not address the bottleneck; managers should first examine the inpatient capacity and the admission rules that decide who reaches it.

\paragraph{Channel design and admission priority.} Within the tested ranges, smaller online shares for both patient groups are preferred when the measures are weighed together, because an online referral enters the admission queue behind in-person referrals and lengthens the tourist wait. Channel expansion should therefore be judged together with the admission rules it interacts with, in line with evidence that the effect of online access depends on its setting \citep{KammrathBetancor2025OnlineBooking}. The same caution applies to priority. Tourists face waiting costs that locals do not face to the same extent, and under the case study's priority order, which ranks the tourist wait first, keeping them ahead in the admission queue is the setting the screening supports. The screening measures only the tourist side of that rule, and how much priority can be given without shifting excessive delay onto local patients remains a management judgment.

\paragraph{Appointment timing.} Within the tested range, a shorter slot interval lowers the system wait and emergencies and raises throughput and utilization, but it also raises early abandonment and places more work on specialists. It can be changed without adding beds or physicians, provided its effects on specialist workload and abandonment are monitored. It does not remove an inpatient bottleneck: in the screening the slot interval has no retained effect on the medical-tourist admission waiting time, so patients who move through consultation faster still wait for the same constrained beds.

\paragraph{Implications of model structure.} The DES-only version appears better on some measures, with no abandonment, fewer emergency escalations, and more recovered patients, because the corresponding behavioral pathways are absent from it. A simpler model can therefore understate operational risks that matter to managers. The value of the hybrid model is that the consequences of patient decisions appear in the same system in which managers compare policies.

\section{Conclusion}\label{sec:conclusion}

Hospitals that serve medical tourists alongside local patients face more than an appointment-scheduling problem: the two groups share specialists, beds, and service channels, and patient decisions change the demand the system sees. This study represents both in a hybrid agent-based and discrete-event simulation.

Bed capacity is the most consistent driver of inpatient waiting, while specialist capacity plays a smaller role in the queueing outcomes; channel design, patient-priority rules, and appointment timing also matter, but their effects depend on how they are combined. In the replicated comparison, the hybrid model produces a mean inpatient admission waiting time of 14.8 days against 25.8 days in the DES-only model, about 42\% lower, and the gap persists under both robustness conditions. That difference is the combined contribution of the behavioral mechanisms in the model as implemented, not the effect of any one of them. A DES-only representation also reports more favorable abandonment, emergency, and throughput figures because it omits abandonment and adherence-driven escalation.

When inpatient capacity is shared across specialties and patient groups, managers should examine bed capacity and the bed-scheduling rules before relying on additional specialist capacity or tighter scheduling, with channel design and patient priority as supporting decisions.

The results come from one hospital, and validation rests on case grounding and face validity, not on predictive back-testing. The screening is directional and rests on a single replicate per design point. The comparison covers three operating points and does not separate the behavioral mechanisms, and the bed-sharing routine used in both model versions does not implement the department's rule. The behavioral inputs are held at their elicited values, the 10-day warm-up is not tuned against the equilibration time of the admission queue, and external shocks lie outside the model. Future work includes implementing the intended bed-sharing rule, varying the behavioral inputs, separating the behavioral mechanisms, replicating the screening design, and testing the framework in other hospitals.

When local and international patients share hospital capacity, performance depends on patient behavior as well as on capacity, and a model that represents both gives managers a clearer basis for deciding where operational changes matter most.

\section*{Statements and Declarations}

\subsection*{Funding}

Not applicable.

\subsection*{Competing interests}

The authors declare no competing interests.

\subsection*{Ethics approval and consent to participate}

This study uses operational process information and expert input obtained from physicians and staff at Takhte Jamshid Hospital and Mehr Clinic during the first author's graduate research at Amirkabir University of Technology. No patients participated, and no patient-level or identifiable information was collected; the article presents only aggregate operational assumptions and simulation outputs. Formal ethics approval was therefore not required for this type of study.

\subsection*{Consent for publication}

Not applicable.

\subsection*{Materials availability}

Not applicable.

\subsection*{Data availability}

This article reports aggregate operational assumptions together with the simulation outputs that follow from case-specific operational information. The underlying hospital operational information is not publicly available for confidentiality reasons and is not released with this article.

\subsection*{Code availability}

The hybrid AnyLogic simulation model (\texttt{IPD27\_hybrid.alp}) is released together with its configuration database in a public repository at \url{https://github.com/MelikaBaghi/IPD-medical-tourism-simulation}. The archived version carries the permanent DOI \url{https://doi.org/10.5281/zenodo.22646799}, and an interactive version runs in a web browser on AnyLogic Cloud (\url{https://cloud.anylogic.com/model/ddca6c4e-7a3b-4015-bd9f-fb8d02cb30be}). The case-specific hospital operational information that sets the model's parameters is held separately and is not redistributed with the model, for the confidentiality reason noted under data availability above.

\subsection*{Use of artificial intelligence}

Documented in Section~\ref{subsec:llm}.

\subsection*{Author contributions}

M.B. built the simulation model, ran the experimental analysis, and drafted the manuscript. H.M. supervised the work, shaped the methodological framing, and revised and edited the manuscript. Both authors have read the final manuscript and approved it.

\bibliography{references}

\end{document}